\documentclass[%
reprint,
 amsmath,amssymb,
 aps,
]{revtex4-2}

\usepackage{float}
\usepackage{slashed}
\usepackage{xcolor}
\usepackage{amsmath}
\usepackage{subcaption}
\usepackage[english]{babel}
\usepackage{lineno, blindtext}
\usepackage{tikz}
\usepackage{adjustbox}
\usepackage{rotating}
\usepackage{rotfloat}
\usepackage{verbatim}
\usepackage[utf8]{inputenc}
\usepackage{graphicx}
\usepackage{dcolumn}
\usepackage{bm}
\usepackage[super]{nth}
\usepackage{lipsum}

\newcommand\Tdiag[4]{%
    \multicolumn{1}{|p{#2}|}{\hskip-\tabcolsep
    \begin{tikzpicture}[%
                baseline={(0,-.25\baselineskip)},
                every node/.style={outer sep=0pt,inner sep=#1}]
    \node[minimum width={#2+1\tabcolsep-\pgflinewidth},
        minimum height=2\baselineskip-\pgflinewidth+\extrarowheight,
        use as bounding box] (box) {};
    \draw[line cap=round] (box.north west) -- (box.south east);
    \node[anchor=south west,text width=.75*#2,align=left] at (box.south west) {#3};
    \node[anchor=north east,text width=.75*#2,align=right] at (box.north east) {#4};
\end{tikzpicture}\hskip-\tabcolsep}}

\usepackage{hyperref}
\usepackage{xcolor}
\hypersetup{
  colorlinks   = true, 
  urlcolor     = red, 
  linkcolor    = blue, 
  citecolor   = blue 
}

\usepackage{threeparttable}
\begin{document}

\preprint{APS/123-QED}

\title{Search for the production of dark matter candidates in association with heavy dimuon resonance using the CMS open data for pp collisions at $\sqrt{s}$ = 8~TeV}



\author{S. Elgammal}
 \altaffiliation[sherif.elgammal@bue.edu.eg]{}
\affiliation{%
 Centre for theoretical physics, The British University in Egypt, Cairo  
}%
\author{M. Louka}%
\affiliation{%
University of Bari and INFN, Bari (Italy)
}

\author{A. Y. Ellithi and M. T. Hussein}
\affiliation{%
 Physics Department, Faculty of Science, Cairo University. 
}%


\date{\today}

\begin{abstract}
{In this work, we present a search for the possible production of Dark Matter particles at the Large Hadron Collider alongside a new hypothetical gauge boson denoted by Z$^{\prime}$, which is governed by a model called Mono-Z$^{\prime}$. The topology of the studied events is dimuons plus large missing transverse momentum. 
The analyzed data were the CMS open data samples collected by the CMS detector, in addition 
to the CMS open Monte Carlo samples, for the proton-proton collisions at 8 TeV centre of mass energy during 2012, which correspond to an integrated luminosity of 11.6 fb$^{-1}$. 
Two benchmarks scenarios were used for interpreting the data, the Dark Higgs scenario as a simplified scenario of the Mono-Z$^{\prime}$ model and the effective field theory formalism of the same model. 
No evidence for the existence of dark matter candidates was found. Consequently, 95$\%$ confidence level limits were set on the masses of the Z$^{\prime}$ and the cutoff scale of the effective field theory.}
\vspace{0.75cm}
\end{abstract}

\maketitle






\section{Introduction}
\label{sec:intro}
The presence of Dark Matter (DM), is one of the most possible explanations for lot of astrophysical observations \cite{R6, R7, R8, R9, R11, R1010}. Results that was reported by the Planck mission  in \cite{planck} indicates that DM represents around 27\% of the observed universe's mass, according to the $\Lambda$CDM model.\\

One of the possible candidates of DM particles are the Weakly Interacting Massive Particles WIMPs, these hypothetical particles are assumed to be weakly-interacting with the baryonic matter. 
Consequently, the strategy of the detection of such particles at the particle colliders is inferred by measuring the momentum imbalance in an event (or missing energy), as DM particles can cross a detector's material without significant interactions. This methodology is applied for the search for DM by the CMS and ATLAS collaborations at the LHC.\\ 

The search for the possible production of DM particles, at the LHC, was performed using events containing a visible particle which will act as a candle, this visible one could be initial or final state radiation (i.e. photon or gluon) or Standard Model (SM) gauge bosons (W/Z) plus large missing transverse momentum. 
This way of search is known in the literature as Mono-X strategy, where X acts as a visible particle that recoil against the dark sector particles.
This idea was applied firstly for the search of DM candidates produced alongside the SM particles at the LHC, as Mono-(W/Z/jets) which were studied in \cite{R35,R45055}, Mono-$\gamma$ and Mono-Higgs results were reported in \cite{photon} and \cite{R36} respectively. 
The same methodology has been extended to include searches for DM alongside new hypothetical, Beyond the Standard Model (BSM), particles \cite{R12}, such as Mono-Z$^{\prime}$ in our present study \cite{R1, ATLAS8}. 
The new Z$^{\prime}$ gauge boson is raised from a lot of extensions to the SM, it is a heavy neutral boson predicted by some of beyond the SM theories \cite{R13}.

The results presented in this paper are complementary to those from the ATLAS collaboration performed at 13 TeV centre of mass energy, that was reported in \cite{R37}. They studied two simplified scenarios of the Mono-Z$^{\prime}$ model considering the hadronic decay of Z$^{\prime}$, while in our present work we study the Effective Field Theory (EFT) formalism of the Mono-Z$^{\prime}$ model in addition to the simplified scenario considering the muonic decay of Z$^{\prime}$. 

In the present work, we used the CMS open Monte Carlo samples and the open CMS real experimental data samples provided by the CMS open data project \cite{R21}, for pp collision at 8 TeV centre of mass energy and corresponding to a total integrated luminosity of $11.6$ fb$^{-1}$.
The status, performance and opportunities for the use of the open CMS data were reported in \cite{R3,Ropendata}.

In the next section, we introduce the theoretical model, in section \ref{section:CMS}, we mention in brief the CMS detector and the reconstruction techniques. The data samples are introduced in section \ref{section:MCandDat}, including the MC simulation of the model signals, the MC simulation of the SM backgrounds and the CMS data sample. The method for the estimation of the contribution of each of the SM background channels is introduced in section \ref{section:Backgrounds}. The search strategy and the selection criteria are discussed in section \ref{section:AnSelection}, followed by the estimation of the total uncertainty in section \ref{section:Uncertainties}, and finally the results including the statistical interpretation and the exclusion limits are introduced in section \ref{section:Results}.

\section{The Mono-Z$^{\prime}$ model}
\label{section:model}
\begin{figure} [h!]
    \centering
    \subfloat[\centering DH]{{\includegraphics[width=5.5cm]{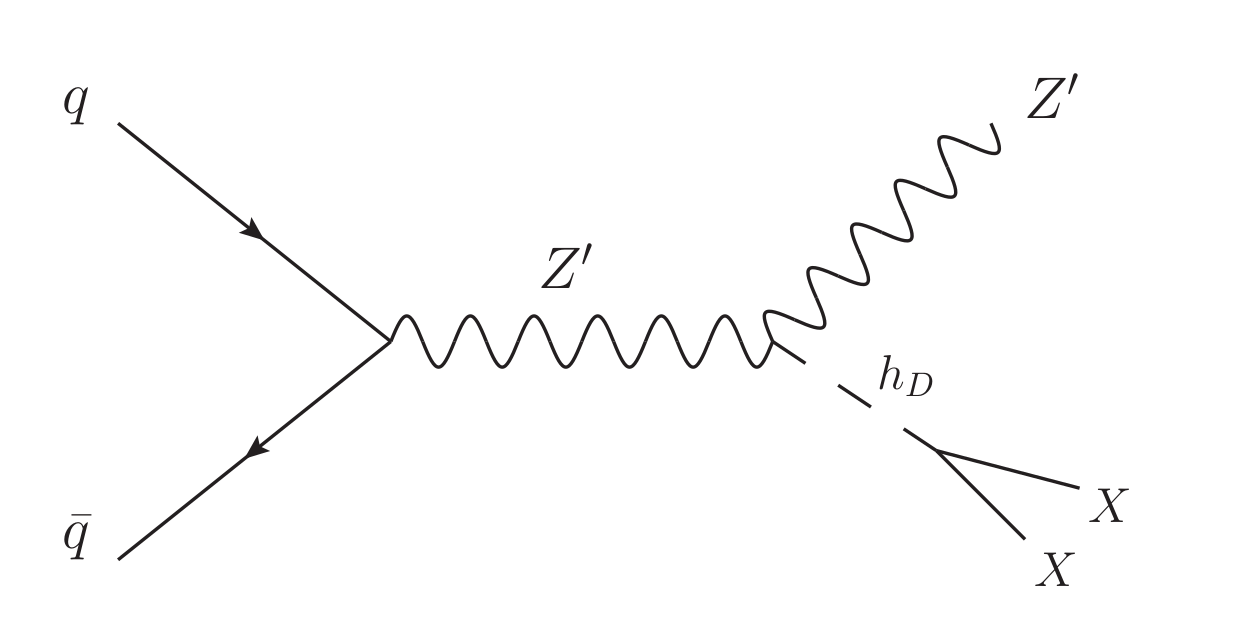} }}%
    \qquad
    \subfloat[\centering EFT]{{\includegraphics[width=5.5cm]{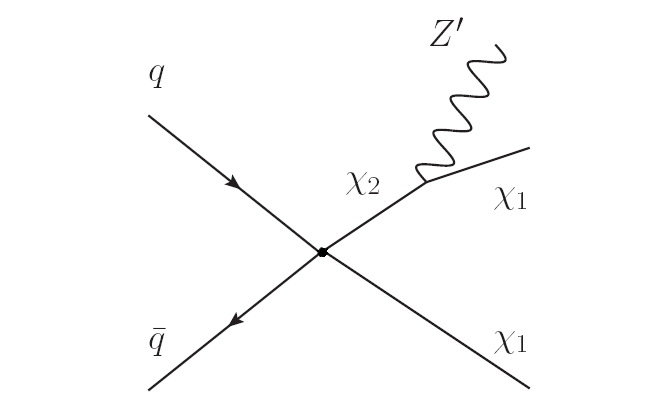} }}%
    \qquad
    \caption{The Feynman diagrams of the two scenarios assumed;(a) dark higgs, (b) light vector with inelastic effective field theory coupling.}%
    \label{figure:fig1}%
\end{figure}

The model introduced in \cite{R1} assumes the production of DM candidate particles alongside the new Z$^{\prime}$ boson. We consider two scenarios, the simplified model, at which the SM fields interact with the DM fields via the mediator vector boson denoted by Z$^{\prime}$, which is the Dark Higgs DH scenario. The second scenario consider the effective coupling between the SM and the DM fields, the Feynman diagrams for the mentioned processes are given in figure \ref{figure:fig1}.

In the DH scenario, Z$^{\prime}$ is produced through $q\bar{q}$ annihilation process in pp collisions, then Z$^{\prime}$ radiates a dark sector Higgs ($h_{D}$) that decays into a pair of final state DM particles ($\chi \bar{\chi}$), the assumptions of the masses followed in this analysis, such that the mass of the dark Higgs is equal to the mass of Z$^{\prime}$, the set which was referred to as "heavy dark sector" introduced in table \ref{table:tab1}. The signal region, for this set, is more shifted from the background dominant region.\\
\begin{table} [h!]    
\centering
\begin {tabular} {ll}
\hline
\hline
Scenario & \hspace{0.8cm} Masses assumptions \\
\hline
\\
    Heavy dark sector & \hspace{0.8cm}  $M_{h_{D}} =$ 
    $\begin{cases}
        125~\text{GeV}, & M_{Z'} < 125~\text{GeV} \\
        M_{Z'}, & M_{h_{D}} > 125~\text{GeV}.
  \end{cases}$\\
 & \\
\hline
\hline
\end {tabular}
\caption{The assumptions of the masses of the particles produced following the DH scenario, the heavy dark sector which is introduced in \cite{R1}.}
\label{table:tab1}
\end{table}
The second scenario, which is known as the EFT scenario reduces the interactions between the DM particles and the SM fields down to contact interaction as given in the following interaction term \cite{R1},
$$ \frac{1}{2\Lambda^{2}}\bar{q}\gamma^{\mu}q(\bar{\chi}_{2}\gamma^{\mu}\gamma^{5}\chi_{1}~+~\bar{\chi}_{1}\gamma^{\mu}\gamma^{5}\chi_{2}).$$

The Feynman diagram, which illustrates this process, is shown in figure \ref{figure:fig1}(b). 
In this scenario, the interaction between the SM fields and the DM fields is introduced by means of the effective coupling between them, the mediator is assumed to be a very heavy (at the TeV scale), hence the interaction approaches a contact interaction. There are two assumed dark states ($\chi_{1}$ and $\chi_{2}$), where $\chi_{2}$ is heavy compared to the mass of Z$^{\prime}$, and the splitting between the the two dark states is enough so that, $\chi_{2}$ can decay into $\chi_{1} + Z^{\prime}$, and $\chi_{1}$ is a final state stable dark fermion. 
The sets of masses followed in this case are given in equation \ref{lv1}, the cross section increases with the lower dark matter masses ($M_{\chi_{1}}$) and this benchmark is optimised for relatively light dark matter mass, where the LHC sensitivity is better \cite{R1}. For heavier DM masses, applying a benchmark for the EFT such that the mass of $\chi_{2}$ is twice the mass of $Z^{\prime}$ while the mass of $\chi_{1}$ is half it's mass, in this case we lose one order of magnitude in the production cross section, hence, this benchmark is not suitable for our current analysis.
\begin{equation}
     M_{\chi_{2}} =  M_{\chi_{1}} + M_{Z^{\prime}} + 25~\text{GeV},~ M_{\chi_{1}} = 5~\text{GeV}.
    \label{lv1}
\end{equation}

The free parameters of the simplified model are; the mass of the mediator $Z^{\prime}$ boson, the masses of the the $M_{h_{D}}$, the coupling between the mediator and the SM fields (particularly quarks) ${\fontfamily{qcs}\selectfont{g}_{SM}}$, in addition to the coupling between the mediator and the DM field ${\fontfamily{qcs}\selectfont{g}_{DM}}$. 

The signature of the Mono-Z$^{\prime}$ process is dilepton or dijet, as products of the Z$^{\prime}$ decaying process, in addition to a missing transverse momentum belongs to the DM candidates. The events analysed in this work have the signature $ \mu^{+}\mu^{-} +\slashed{p}_{T}$.

For the DH scenario, the Mono-Z$^{\prime}$ samples were simulated for mediator masses between 150 GeV and 700 GeV (higher values of the mediator have not been studied due to the dramatic drop in their cross section measurements, consequently the statistical analysis could not be done properly), and from 1 to 200 GeV for the mass of DM; with the $g_{DM}$ coupling value set to $g_{DM}$ = 1 \cite{R1, R37}. Following the experimental constraints from dijet resonance searches by CMS collaboration at the LHC with 8 TeV centre of mass energy \cite{RefCoupling}, in particular those for the mediator mass range below about 500 GeV studied in this analysis, the $g_{SM}$ coupling value was set to 0.25.
The cross section measurements multiplied by the $ Z^{\prime} \rightarrow \mu^{+}\mu^{-} +\slashed{p}_{T}$ branching ratios for the DH scenario are listed in tables \ref{table:tabchi}, they are varied with the change of both of the mediator and dark higgs masses, and do not depend on the choice of dark matter mass.

For the EFT, the cutoff scale ($\Lambda$) is the main parameter of the model, it is the energy scale beyond which the approach becomes invalid. 
The EFT production cross section measurements times branching ratios as a function of the scenario cutoff scale of the EFT($\Lambda$), for a fixed mass point of Z$^{\prime}$ ($M_{Z^{\prime}}$ = 450 GeV) and centre of mass energy $\sqrt{s} = 8$ TeV, are listed in
table \ref{table:EFT}.

All of these cross section measurements times branching ratios are calculated with the use of MadGraph5 aMC@NLO v2.6.7 \cite{R33} at next-to-leading order, regarding pp collisions at the LHC with 8 TeV centre of mass energy.
The decay widths for each of the Z$^{\prime}$ and the $h_{D}$ for the DH scenario, or Z$^{\prime}$ and $\chi_{2}$ for the EFT case, that increase with the particles' masses, are calculated at each mass value within MadGraph5 \cite{dwidth}. For the $Z^{\prime}$ resonance, the decay width ranges from 4.47 to 24.28 GeV over the scanned range of masses i.e. between 150 and 700 GeV, with a width of 8.97 GeV at the used benchmark mass point for the EFT interpretation.


\begin{table} [p]
\begin{sidewaystable} [H]
\centering
\begin{tabular}{|c||c|c|c|c|c|c|c|c|c|c|c|}
\hline
\Tdiag{.01em}{1.6cm}{$M_{\chi}$\tiny{(GeV)}}{$M_{Z'}$\tiny{(GeV)}}  & 150 & 200 & 300 & 325 & 350 & 375 & 400 & 425 & 450 & 475 & 500 \\
\hline
\hline
1  & $7.10\times10^{-02}$ & $2.36\times10^{-02}$ & $0.438\times10^{-02}$ & $0.305\times10^{-02}$  & $0.2107\times10^{-02}$ & $0.144\times10^{-02}$ & $0.1036\times10^{-02}$ & $0.764\times10^{-03}$ & $0.568\times10^{-03}$ & $0.428\times10^{-03}$ & $0.328\times10^{-03}$ \\
\hline
5  & $7.08\times10^{-02}$ & $2.37\times10^{-02}$ & $0.437\times10^{-02}$ & $0.306\times10^{-02}$ & $0.210\times10^{-02}$ & $0.144\times10^{-02}$ & $0.104\times10^{-02}$ & $0.763\times10^{-03}$ & $0.569\times10^{-03}$ & $0.427\times10^{-03}$ & $0.3283\times10^{-03}$\\
\hline
10 & $7.10\times10^{-02}$ & $2.36\times10^{-02}$ & $0.437\times10^{-02}$ & $0.305\times10^{-02}$ & $0.211\times10^{-02}$ & $0.145\times10^{-02}$ & $0.104\times10^{-02}$ & $0.763\times10^{-03}$ & $0.569\times10^{-03}$ & $0.429\times10^{-03}$ & $0.328\times10^{-03}$\\
\hline
25 & $7.10\times10^{-02}$ & $2.358\times10^{-02}$ & $0.437\times10^{-02}$ & $0.305\times10^{-02}$ & $0.211\times10^{-02}$ & $0.144\times10^{-02}$ & $0.1035\times10^{-02}$ & $0.763\times10^{-03}$ & $0.568\times10^{-03}$ & $0.429\times10^{-03}$ & $0.329\times10^{-03}$\\
\hline
50 & $7.13\times10^{-02}$ & $2.36\times10^{-02}$ & $0.437\times10^{-02}$ & $0.306\times10^{-02}$  & $0.210\times10^{-02}$ & $0.144\times10^{-02}$ & $0.1038\times10^{-02}$ & $0.756\times10^{-03}$  & $0.567\times10^{-03}$ & $0.429\times10^{-03}$ & $0.328\times10^{-03}$\\
\hline
75 & $16.40\times10^{-07}$  & $2.36\times10^{-02}$ & $0.436\times10^{-02}$ & $0.305\times10^{-02}$ & $0.209\times10^{-02}$ & $0.144\times10^{-02}$ & $0.104\times10^{-02}$  & $0.763\times10^{-03}$ & $0.568\times10^{-03}$ & $0.429\times10^{-03}$ & $0.328\times10^{-03}$ \\
\hline
100 & $8.98 \times10^{-07}$ & $5.43\times10^{-07}$ & $0.436\times10^{-02}$ & $0.3052\times10^{-02}$ & $0.211\times10^{-02}$ & $0.144\times10^{-02}$ & $0.1039\times10^{-02}$ & $0.764\times10^{-03}$ & $0.568\times10^{-03}$ & $0.428\times10^{-03}$ & $0.326\times10^{-03}$ \\
\hline
125 & $1.54\times10^{-07}$ & $5.01\times 10^{-07}$ & $0.437\times10^{-02}$ & $0.3049\times10^{-02}$ & $0.209\times10^{-02}$ & $0.144\times10^{-02}$ & $0.104\times10^{-02}$ & $0.758\times10^{-03}$ & $0.567\times10^{-03}$ & $0.429\times10^{-03}$ & $0.327\times10^{-03}$ \\
\hline 
150 & $4.09 \times10^{-08}$ & $1.0 \times10^{-07}$ & $1.00\times10^{-07}$ & $0.3047\times10^{-02}$ & $0.2094\times10^{-02}$ & $0.144\times10^{-02}$ & $0.1037\times10^{-02}$ & $0.758\times10^{-03}$ & $0.567\times10^{-03}$ & $0.427\times10^{-03}$ & $0.326\times10^{-03}$ \\
\hline
175 & $1.3\times10^{-08}$ & $3.05\times10^{-08}$ & $1.753 \times10^{-07}$ & $3.766 \times10^{-07}$ & $0.48\times10^{-07}$ & $0.1436\times10^{-02}$ & $0.1035\times10^{-02}$ & $0.756\times10^{-03}$ & $0.556\times10^{-03}$ & $0.427\times10^{-03}$ & $0.326\times10^{-03}$ \\
\hline
200 & $5.25 \times10^{-09}$ & $1.0 \times10^{-08}$ & $4.36 \times10^{-08}$ & $6.56 \times10^{-08}$ & $1.04 \times10^{-07}$ & $2.08 \times10^{-07}$ & $0.239\times10^{-07}$ & $0.76\times10^{-03}$ & $0.566\times10^{-03}$ & $0.427\times10^{-03}$ & $0.325\times10^{-03}$ \\
\hline
\end {tabular}

\caption{The cross section measurements times branching ratios in pb, for the DH scenario, calculated at different values of the mass of $M_{Z'}$ and the mass of the DM particle $M_{\chi}$, assuming the heavy dark sector set of masses, the couplings constants are taken to be $g_{SM} = 0.25,~g_{DM} = 1.0$, at 8 TeV centre of mass energy.}
\label{table:tabchi}
\end{sidewaystable}
\end{table}
\begin{table} [h!]
\centering
\begin{tabular}{|c|c|}
\hline
    $\Lambda (TeV)$ & $\sigma~\times$ BR~~(pb)  \\
\hline
\hline
    1.0 & 0.0704  \\
\hline
    1.5 & 0.0139 \\
\hline
    2.0 & 0.0044  \\
\hline
    2.5 & 0.0018 \\
\hline
     3.0& 0.00087  \\
\hline
     3.5& 0.00047  \\
\hline
     4.0& 0.000275  \\
\hline
     5.0& 0.0001122  \\
\hline
\end{tabular}
\caption{The EFT production cross section measurements times branching ratios as a function of the scenario cutoff scale of the EFT($\Lambda$), for a fixed mass point of Z$^{\prime}$ ($M_{Z^{\prime}}$ = 450 GeV) and centre of mass energy $\sqrt{s} = 8$ TeV.}
\label{table:EFT}
\end{table}

\section{The CMS detector and reconstruction techniques}
\label{section:CMS}
The Compact Muon Solenoid CMS is a general purposed particles detector that has been located at one of the four collision points of the LHC, that provides a facility to search for new physics BSM at the TeV scale,  a description of the detector and its performance with details can be found in \cite{R17,R29}. The CMS can be divided into five main layers, the first layer surrounding the beam pipe is the tracker, the electromagnetic calorimeter, the hadron calorimeter that plays an essential rule for the reconstruction of the missing transverse momentum, which is the signature of neutrinos or other exotic weakly interacting particles e.g. DM candidates in our case.\\
The superconducting magnet, provided by the installed solenoid, is the \nth{4} layer laying between the hadron calorimeter and the muon system. The last layer that envelope the detector is the muon system, the barrel part of the muon system covers the pseudorapidity range $|\eta| < 1.2$ while the end-caps cover the range $1.2 < |\eta| < 2.4$.

As we concern with events containing dimuon and missing transverse momentum, we refer to the identification and reconstruction of muons described in \cite{R18,R40}, and the reconstruction of the missing transverse momentum $\slashed{p}_{T}$ described in \cite{R19}.
Certain corrections must be applied while reconstructing the $\slashed{p}_{T}$, in order to get much better data to MC agreement for all distributions related to missing transverse momentum as mentioned in \cite{R45}, as the negative sum of the momenta of the particle flow (PF) objects, these corrections account for some factors affecting the values of the reconstructed PF ${p}_{T}$, and hence, have a direct impact on the calculations. These factor are; inefficiencies in the tracker, the $p_{T}$ threshold and the energy threshold in the tracker and calorimeters respectively, and the non-linear response of the calorimeters for hadrons \cite{R45}. The formula, that is used to calculate the $\slashed{p}_{T}$ after applying the mentioned corrections, is given in equation \ref{pt}.
\begin{equation}
    \vec{\slashed{p}}_{T}^{~\text{corr}} = \vec{\slashed{p}}_{T} - \sum_{jets} (\vec{p}_{T jet}^{~\text{corr}} - \vec{p}_{T jet})
    \label{pt}
\end{equation} 
where $\vec{\slashed{p}}_{T}^{~\text{corr}}$ and $\vec{p}_{T jet}^{~\text{corr}}$ refer to the corrected values considering the inefficiencies mentioned above. the equation indicates that the difference between the PF jet  $\vec{p}_{T}$ after and before applying the correction, must be subtracted from the reconstructed  $\vec{\slashed{p}}_{T}$ in order to make correct balancing.
\section{Data and Monte Carlo samples}
\label{section:MCandDat}
\subsection{The samples of the model signals}
We used the matrix element event generator \text{MadGraph5\_aMC@NLO}~v2.6.7 \cite{R33} for the generation of events processes. The generation process and the calculation of the signal cross section had been done at next-to-leading order (NLO), based on the Universal FeynRules Output UFO provided by the authors of \cite{R1}. This generator has been interfaced with \text{P{\footnotesize YTHIA}} v.6.4.26 for the modeling of the parton showers \cite{R34}, the NNPDF2.3QED NLO set for the parton distribution functions (PDF) \cite{PDFsignal} has been considered, which is available via the LHAPDF6 library \cite{LHAPDF6}.
The full simulation for the CMS detector and the reconstruction process have been done within the standard CMSSW frame work, the simulation of the detector response to the particles has been performed with \text{G{\footnotesize EANT4}} \cite{geant4}, and the reconstruction process is done with the the release \textit{CMSSW\_5\_3\_32} \cite{CMSSWversion} which is the relevant version to the CMS open data. The reconstruction process is done regarding the requirements of the experimental set up of the CMS detector during the LHC run-I at $\sqrt{s}=8$ TeV, this requirements will be introduced and discussed in the preselection of events at section \ref{section:AnSelection}.

\subsection{The SM backgrounds samples}

There are a lot of SM processes which decay to dimuon in the final state plus missing transverse momentum attributed to SM neutrinos, the muons basically come from the decay of the SM gauge bosons. We have considered the most significant channels for building the SM background which are; the top quark pairs production ($t \bar{t}$), the electroweak diboson production (WW, WZ, ZZ) and the Drell-Yan (DY) process. In the presented study, we have used the CMS open MC samples generated by the CMS collaboration to construct the SM backgrounds, these samples are available via the CERN open data portal \cite{portal}, the used samples are listed in table \ref{table:tab3}. The processes for the production of a single top quark alongside W boson are ignored in this analysis, basically because they are not included in the list of CMS open MC samples. 

The cross section measurements of the MC samples used in this analysis, at next-to-leading order (NLO) or next-to-next-to-leading order (NNLO), are indicated in table \ref{table:tab3}. 
The cross section of the $t\bar{t}$ process was calculated using the parton-level Monte Carlo program MCFM \cite{ttbar}, and the rest of the samples' cross sections were taken from \cite{R3}. 
The Dell-Yan and ZZ processes have been generated with \text{P{\footnotesize OWHEG}B{\footnotesize ox}} v1.0 MC program \cite{powheg1,powheg2} interfaced to \text{Pythia v.6.4.26} parton shower model \cite{R34}.
The rest of other MC background samples were modeled with \text{MadGraph5\_aMC@NLO}~v5.1.3.30 \cite{MG2} event generator interfaced to \text{Pythia v.6.4.26}. 
The CT10 parton distribution functions (PDFs) \cite{PDF} and the Z2* PYTHIA6 tune \cite{Z21,Z22} have been used.

The effect of the pile-up has been simulated by overlying the generated MC events with a minimum bias, finally the response of the CMS detector to the particles is simulated by \text{G{\footnotesize EANT4}} \cite{geant4}.

\begin{table*}[]
\centering
\begin {tabular} {|l|l|c|c|}
\hline
Channel & Data-Set file & ${\Large \sigma} \times~\text{BR}$ ~(\text{pb}) & Order\\
\hline
\hline
$DY \rightarrow \mu\bar{\mu}$  & DYToMuMu\_M-20\_CT10\_TuneZ2star\_v2\_8TeV. \cite{R22}  & 1916 \cite{R3}& NNLO\\
\hline
$t\bar{t}$ + jets   & TTJets\_FullLeptMGDecays\_8TeV. \cite{R23} & 23.89 \cite{ttbar}& NLO\\
\hline
WW + jets  & WWJetsTo2L2Nu\_TuneZ2star\_8TeV. \cite{R24} & 5.8 \cite{R3}& NLO\\
\hline
WZ + jets   & WZJetsTo3LNu\_8TeV\_TuneZ2Star. \cite{R25} &1.1 \cite{R3}& NNLO\\
\hline
$ZZ\rightarrow 4\mu$   & ZZTo4mu\_8TeV. \cite{R26} & 0.077 \cite{R3}& NLO\\
\hline
\end {tabular}
\caption{The Monte Carlo samples used to build the SM background with the corresponding cross sections times the branching ratios. the data-set name for each sample - as appear on the open data portal - are also stated, these samples are generated for pp collision at $\sqrt{s}= 8$ TeV.}
\label{table:tab3}
\end{table*}

 
\begin{table*}[]
\centering
\label{ tab-marks }
\begin {tabular} {|c|l|c|}
\hline
Run & Data Set & $\mathcal{L}$ ({fb}$^{-1}$) \\
\hline
\hline
Run-B~ & SingleMu/Run2012B-22Jan2013-v1/AOD.\cite{R27}  &  \\ 
   &   & 11.6 \cite{Ropendata}\\
Run-C~ & SingleMu/Run2012C-22Jan2013-v1/AOD.\cite{R28}  & \\
\hline
\end {tabular}
\caption{The used CMS-2012 open data samples collected by CMS experiment during  
the LHC run-{\footnotesize I} at 8 TeV centre of mass energy, and the corresponding 
integrated luminosity.}
\label {table:tab4}
\end{table*}

\subsection{The CMS open data samples}
\label{section:CMSopenData}
The CMS open experimental data samples, used in this analysis, are based on pp collision at $\sqrt{s}$ = 8 TeV during the LHC run-{\footnotesize I} and recorded by the CMS detector in 2012.
We used the two open data runs (run-{\footnotesize B} and run-{\footnotesize C}) corresponding to a total integrated luminosity of 11.6~fb$^{-1}$ \cite{Ropendata}, more detail about these runs are listed in table \ref{table:tab4}. 
The data were triggered by the high level trigger HLT$\_$Mu40$\_$eta2p1 which is a single muon trigger. 
This trigger was unprescaled for the full 2012 data-set and aim to collect events with at least one muon candidate within $|\eta| <$ 2.1 and $\text{p}_{\text{T}} > 40$ GeV. 
The efficiency of this trigger varies as a function of $\eta$, resulting in an efficiency for triggering on a dimuon system that varies between 97\% and 100\% \cite{zprime}.
The events have been taken from the list of the validated runs (known as the \text{good runs list}), for the primary sets of 2012 data provided by the open data project \cite{R39}, at which all the CMS sub-detectors were working stably. 
The samples, their data-sets names and the corresponding integrated luminosity ($\mathcal{L}$) are listed in table \ref{table:tab4}.

\section{Backgrounds estimation}
\label{section:Backgrounds}

There are many background processes that include dimuon in the final state plus missing transverse momentum, and can mimic with our events topology in our search for new physics. 
The first type is the SM processes produced during pp collisions, the second is the jets contamination 
and the third is the cosmic muons background.

The contribution of the SM background processes, that are considered in the present study, have been estimated from the Monte Carlo simulations, following the same method applied in the previous search for new resonance within the dimoun events at $\sqrt{s}= 8$ TeV\cite{zprime}. The Monte Carlo sample of the SM backgrounds, which are listed in table \ref{table:tab3}, are normalized to their corresponding cross sections.
The jets background arises from the misidentification of jets as muons, where a jet or multijet pass the muons selection criteria. This kind of backgrounds comes from two processes; W+jet and QCD multijet. 
The contamination of single and multi jets background in data is usually estimated from data using a so called data driven method which is explained in \cite{zprime}.  
It has been founded that the QCD and W+jets contributions are very small above 400 GeV at the dimuon invariant mass spectrum, as estimated in \cite{zprime}, with only 3 events could be misidentified as muons for an integrated luminosity of 20.6 fb$^{-1}$, thus in our case (luminosity = 11.6 fb$^{-1}$) this contribution is expected to be much lower than 3 events and is negligible in the current study.

The last background source comes from the Cosmic muons that cross the detector layers and pass near the interaction point while the operation process, this background can be suppressed by constraining the vertex position and the impact parameter associated with the reconstructed muon. A cut is applied such that the muon's transverse impact parameter, with respect to the primary vertex, must be less than 0.2 cm. For cosmic muons that pass in-time with a collision event, and pass the vertex position and the impact parameter cuts, the 3D angel between each of the reconstructed dimuons, is restricted to be below $\pi - 0.02$ rad. The mentioned cuts are applied in the identification of muons in the 2012-analysis \cite{R41, R32}. After all, it has been founded that the cosmic muons contribution to our background is less than 0.1 events, and can be also neglected \cite{zprime}.
\section{Selection of events}
\label{section:AnSelection}
\subsection{Preselection of events}
\label{section:Preliminary}
The aim of the selection is to pick out events containing dimuon in addition to missing transverse momentum. This selection is divided into two steps; the first one is the preselection which is presented in table \ref{cuts}$(i)$ and the second step is the tight selection introduced in table \ref{cuts}$(ii)$. The detailed definition of these cuts will be explained in this section.

The preselection is a manifestation of the high transverse momentum $(p_{T})$ muon identification introduced in \cite{R41, R32}. It includes cuts related to the trigger requirements (HLT\_Mu40\_eta2p1), the $p_{T}$ threshold of this trigger is 40 GeV within the tracked acceptance ($|\eta| < 2.1$) and the high $p_{T}$ muon ID, that was applied in 2012 data analysis, used for the search for new physics with events containing dimuon resonance \cite{zprime}. In addition we apply some kinematics cuts as the reconstructed transverse momentum of the muon ($p^{\mu}_{T}$) must be greater than $45$ GeV, $|\eta^{\mu}| < 2.1$ and the invariant mass of the dimuon must be above 50 GeV, as we are looking for a resonance in the high mass regime. The preselection cuts are listed in table \ref{cuts}$(i)$. 

\begin{figure}[h]
\centering
\resizebox*{9cm}{!}{\includegraphics{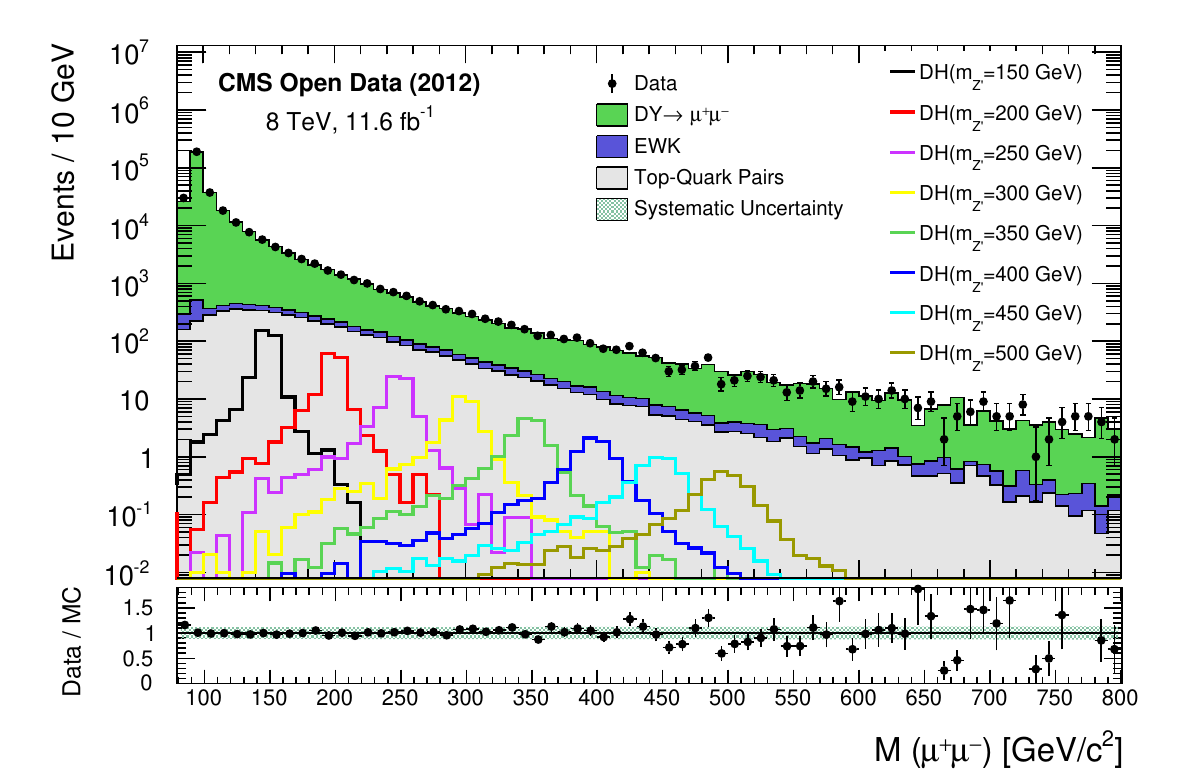}}
\caption{The distribution of the invariant mass of the dimuon, after applying preselection of events (listed in table \ref{cuts}), for the CMS data, the expected SM channels and Z$^{\prime}$ masses generated regarding the DH model. The lower band shows the data-to-simulation ratio with an illustration of the total uncertainty in the estimation of the expected background (shaded region).}
\label{figure:fig3}
\end{figure}
\begin{figure}[h]
\centering
\resizebox*{9cm}{!}{\includegraphics{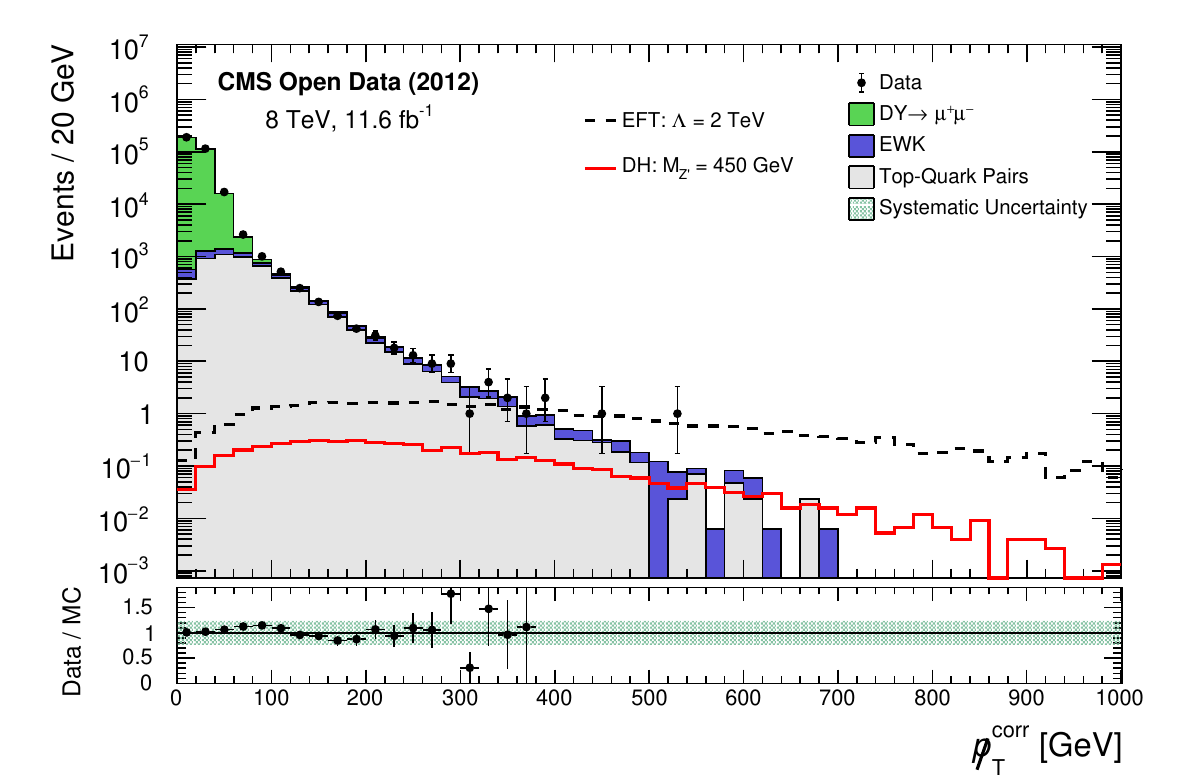}} 
\caption{The distribution of the missing transverse momentum , after the preselection (listed in table \ref{cuts}); for the CMS data, the expected SM backgrounds, a DH model signal produced at $M_{Z^{\prime}} = 450$ GeV, and an EFT model signal produced at $\Lambda = 2$ TeV. The lower band shows the data-to-simulation ratio with an illustration of the total uncertainty in the estimation of the background (shaded region). Signals are normalized to the product of the cross section times the branching ratio of the muonic decay of Z$^{\prime}$.}
\label{figure:fig4}
\end{figure}

The dimuon invariant mass distribution, after the application of the preselection, is shown in figure \ref{figure:fig3}. The CMS data are represented by black dots with vertical bars (accounts for statistical error), the DY background is represented by the green histogram, the grey histogram represents the $t \bar{t} + jet$ background, and the electroweak diboson backgrounds (WW, WZ and ZZ) are added together and represented by the blue histogram. 
Signals attributed to DH  model, at different values of the mediator's mass $M_{Z^{\prime}}$, are overlaid. The lower band shows the ratio between the CMS data and the expected SM background, with an illustration of the total uncertainty related to the prediction of this background processes (which will be discussed in section \ref{section:Uncertainties}). 
In the rest of this paper, all figures follow the same plotting style and keys.
Figure \ref{figure:fig4} shows the distribution of the missing transverse momentum $\slashed{p}^{corr}_{T}$, after the application of the preselection of events, the CMS data, the SM expected backgrounds, a model signal for the DH scenario at $M_{Z^{\prime}} = 450$ GeV, and a model signal for the EFT scenario at $M_{Z^{\prime}} = 450$ GeV and $\Lambda = 2$ TeV are included.

It is clearly noticed, in figures \ref{figure:fig3} and \ref{figure:fig4} , that the CMS data is in a good agreement with the Monte Carlo simulation of the expected SM processes within the total uncertainty. Moreover, the Mono-Z$^{\prime}$ model signals are totally submerged by the SM background, these problem has been successfully handled by applying extra cuts which will be discussed in the following subsection.\\
\subsection{Events selection}
The events selection is a combination between the preselction cuts introduced in table \ref{cuts}$(i)$ and extra tighter cuts, presented in table \ref{cuts}$(ii)$, based on 
four variables. The first variable is related to the invariant mass of the dimuon, at which we restricted the invariant mass of the dimuon to a small range around the mass of the Z$^{\prime}$, such that $(0.9 \times M_{Z^{\prime}}) < M_{\mu^{+}\mu^{-}} < (M_{Z^{\prime}} + 25)$, the aim of this cut is to suppress the Drell-Yan peak. 
The second is the difference in the azimuthal angle between the dimuon and the missing transverse momentum vector ($\Delta\phi_{\mu^{+}\mu^{-},\vec{\slashed{p}}_{T}^{\text{corr}}}$), it has been selected to be greater than 2.6 rad., this cut is optimized to the model signals region.
The Third one is the relative difference between the $p_{T}$ of dimuon and the missing transverse momentum ($|p_{T}^{\mu^{+}\mu^{-}} - \slashed{p}_{T}^{\text{corr}}|/p_{T}^{\mu^{+}\mu^{-}}$), it has been selected to be less than 0.6 which is an optimized cut to the signals region. 
Final, a very tight cut on the value of $\slashed{p}_{T}^{\text{corr}}$ was applied in order to suppress DY, ZZ,  W+jets and the QCD contributions.
\begin{table}[H]
    \centering
    \begin {tabular} {|c|c|c|}
\hline
step & variable & requirements \\
\hline
    \hline
    & Trigger  & HLT\_Mu40\_eta2p1 \\
    & High $p_{T}$ muon ID & \cite{R41, R32}\\
(\romannumeral 1) & $p^{\mu}_{T}$ (GeV) & $>$ 45 \\
    & $\eta^{\mu}$ (rad) & $<$ 2.1 \\
    & $M_{\mu^{+}\mu^{-}}$ (GeV) & $>$ 50 \\
    
\hline
     &Mass window (GeV)& $(0.9{\times} M_{Z^{\prime}})< M_{\mu^{+}\mu^{-}} < (M_{Z^{\prime}}{+}25)$\\
(\romannumeral 2)     &$|p_{T}^{\mu^{+}\mu^{-}} - \slashed{p}_{T}^{\text{corr}}|/p_{T}^{\mu^{+}\mu^{-}}$ & $<$ 0.6  \\
     &$\Delta\phi_{\mu^{+}\mu^{-},\vec{\slashed{p}}_{T}^{\text{corr}}}$ (rad) & $>$ 2.6 \\
     
     &$\slashed{p}_{T}^{\text{corr}}$ (GeV) & $>$ 100 \\
    \hline
    \end{tabular}
    \caption{Summary of the preselection cuts (\romannumeral 1) and the cut-based final events selection (\romannumeral 1 +\romannumeral 2) applied in the analysis.}
    \label{cuts}
\end{table}
We have investigated the efficiency of the selection, we have found that the efficiency of the selection with respect to the DH signal is around 67\% upon the entire scanned range in the $\slashed{p}^{corr}_{T}$ distribution, moreover, regardless the control region which is rich with background events, the efficiency is around 80\% for the DM signal, above a transverse momentum of 200 GeV at the signal region. 
Moreover, most of the SM background events are suppressed by applying the events selection introduced in table \ref{cuts}.
\section{Systematic uncertainties}
\label{section:Uncertainties}
A variety of sources of systematic uncertainties have been considered while interpreting the results. Some sources originate from experimental issues, other sources are theoretical and related to the uncertainty in the Parton Distribution Functions PDF, that were used during the production process of the SM samples. The different sources of the systematic uncertainties, considered in the presented results, are listed in table \ref{table:sources}.

\begin{table}[h!]
    \centering
    \begin{tabular}{l r}
    \hline
    \hline
    feature     & Uncertainty (\%) \\
    \hline
    Luminosity ($\mathcal{L}$) & 2.6 {\footnotesize \cite{Lumi}} \\
    
    $A\times\epsilon$     & 3 {\footnotesize \cite{zprime}}\\
    
    $p_{T}$ resolution & 5 {\footnotesize \cite{zprime}}\\
    
    $p_{T}$ scale & 5 {\footnotesize \cite{zprime}}\\
   
    Unclustered $\slashed{\slashed{p}}_{T}^{~\text{corr}}$ scale & 10 {\footnotesize \cite{R45}} \\
    
    Jet energy scale & 2-10 {\footnotesize \cite{R45}} \\
    
    Jet energy resolution & 6-15 {\footnotesize \cite{R45}} \\
    
    PDF (Drell-Yan) & 4.5 {\footnotesize \cite{zprime}}\\
    
    PDF (ZZ) & 5 {\footnotesize \cite{R450}}\\
    
    PDF (WZ)  & 6 {\footnotesize \cite{R450}}\\ 
    \hline
    \hline
    \end{tabular}
    \caption{Sources of systematic uncertainties considered in the presented analysis, and their values in percentage.}
    \label{table:sources}
\end{table}
\section{Results}
\label{section:Results}
The assumed masses of DM particles are much heavier than the SM neutrinos, specially for the heavy dark sector set of masses applied in this analysis and introduced in section \ref{section:model}. Consequently, the distribution of the missing transverse momentum attributed to any DM signal is expected to be characterized by longer tail than the corresponding distribution from the SM neutrinos, which could be a good discriminator between the two hypotheses. 
For this reason we have applied a shape analysis strategy regarding the $\slashed{p}^{corr}_{T}$ distributions to our study.

\begin{figure}[h]
\centering
  \includegraphics[width=\linewidth]{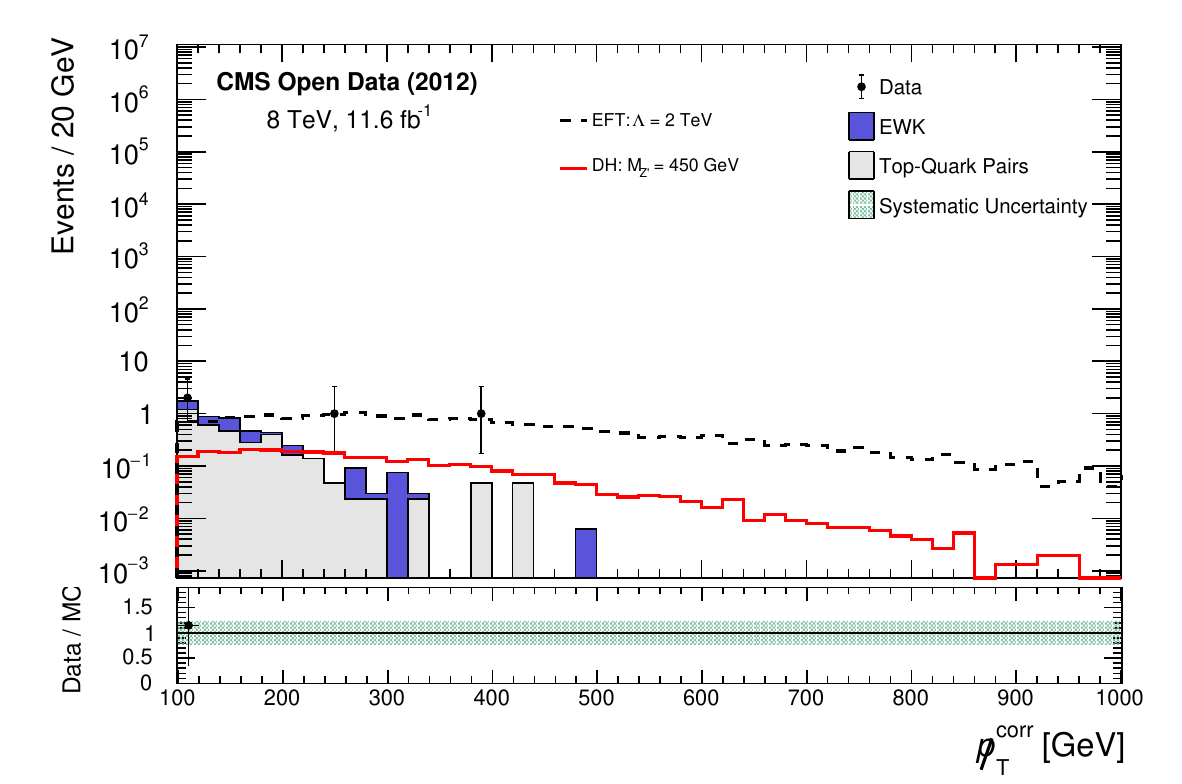}
  \caption{The distribution of the missing transverse momentum , after applying the events selection (listed in table \ref{cuts}); for the CMS data, the expected SM backgrounds, a DH model signal produced at $M_{Z^{\prime}} = 450$ GeV, and an EFT model signal produced at $\Lambda = 2$ TeV. The lower band shows the data-to-simulation ratio with an illustration of the total uncertainty in the estimation of the background (shaded region). Signals are normalized to the product of the cross section times the branching ratio of the muonic decay of Z$^{\prime}$.}
  \label{figure:fig6}
\end{figure}

The $\slashed{p}^{corr}_{T}$ distribution, after applying the criteria of the events selection summarized in table \ref{cuts}, for each of the CMS data, the SM background channels and the mono-Z$^{\prime}$ signals, is shown in figure \ref{figure:fig6}. A significant reduction of the SM backgrounds is achieved by applying the cuts introduced in the events selection, 
moreover the distribution shows that, the CMS data is in a good agreement with the SM simulated processes within the total uncertainty. 
The two stray points in the $\slashed{p}^{corr}_{T}$ distribution (above 250 GeV ), can be explained in terms of the statistical fluctuations due to the leakage of the data points at the high missing transverse momentum regime, we have found that these events are in an agreement with the expected background within 1.43 $\sigma$ significance.

The number of events passing the criteria of the events selection, for each of the CMS data, the SM background channels, a DH signal (at $M_{Z^{\prime}} = 450$ GeV) and an EFT signal (at $M_{Z^{\prime}} = 450$ GeV and $\Lambda = 2$ TeV), are listed in table \ref{table:tab8}, the total uncertainty, including the statistical and systematic components, are also indicated, this summation has been done using a quadrature formula.


\begin{table}[h!]
\centering
\label{ tab-marks }
\begin {tabular} {|l|c|}
\hline
Process & No. of events \\
\hline
\hline
$t\bar{t} + jets$ & $3.5 \pm   2.0$\\
\hline
$WW + jets$ & 1.5 $\pm   1.3$ \\
\hline
$WZ + jets$ & 0.2 $\pm   0.4$ \\
\hline
$ZZ \rightarrow 4\mu$  & $0.001 \pm   0.034$   \\
\hline
Sum Bkgs & 5.1 $\pm  2.6$  \\
\hline
\hline
DH signal & 2.9 $\pm   1.8$   \\
(at $M_{Z^{\prime}}$ = 450 GeV) &  \\ 
\hline
EFT signal & 21.8 $\pm   7.0$   \\
(at $\Lambda$ = 2 TeV \& $M_{Z^{\prime}}$ = 450 GeV) & \\ 
\hline
\hline
Data & 4 \\
\hline
\end {tabular}
\caption{The number of events satisfying the criteria of the events selection, for each SM background, the model signals and the CMS open data; corresponding to a 11.6 fb$^{-1}$ integrated luminosity. The total  Uncertainty, including the statistical and systematic components, is indicated.}
\label{table:tab8}
\end{table}
\subsection{Statistical interpretation}
A statistical test based on the profile likelihood method, with the use of the modified frequentist construction CLs \cite{R58, R59} used in the asymptotic approximation \cite{R2}, has been employed for making statistical interpretation of the results, i.e setting limits on the model. 
The SM background-only hypothesis has been tested against the Mono-Z$^{\prime}$ signal hypothesis.
The signal hypothesis is excluded at a probability value equal to 0.05, corresponding to a significance of 2$\sigma$, the p-value could be defined as the probability that the incompatibility of the data with the SM background-only hypothesis, is a manifestation of the systematic uncertainty. The confidence intervals (fluctuations about the median) are calculated within 1$\sigma$ or 2$\sigma$, corresponding to confidence levels of 68\% or 95\%, respectively.

The likelihood function can be expressed in terms of the product of the Poisson probabilities as the following:- 
$$ \mathcal{L}(\mu,\theta) = \prod_{i=1}^{M} \frac{{(\mu s_{i} + b_{i})}^{n_{i}} ~e^{-(\mu s_{i} + b_{i})} }  {n_{i}~!}\prod_{j=1}^{k} \frac{ u_{j}^{m_{j}} ~e^{-u_{j}} }  {m_{j}~!}, $$
where the first $\pi$-product accounts for the signal strength ($\mu = \sigma / \sigma_{th} $), our parameter of interest and the second $\Pi$-product is for the nuisance parameters ($\theta$), which accounts for the background normalization and the systematic uncertainties. $s_i$ and $b_i$ are number of signal and background events per each bin, respectively, estimated from the simulation, while $u_{j}$ is a function of $\theta$ that gives the expectation value for each bin in the control sample used to constrain this nuisance parameters. Using of shape analysis applied to the $\slashed{p}^{corr}_{T}$ distributions, we have constructed the signal strength as a function of the model free parameters to be constrained, the limits obtained are introduced at the next section.
\subsection{Exclusion limits}

\begin{figure}[h!]
\centering
  \resizebox*{8.cm}{!}{\includegraphics{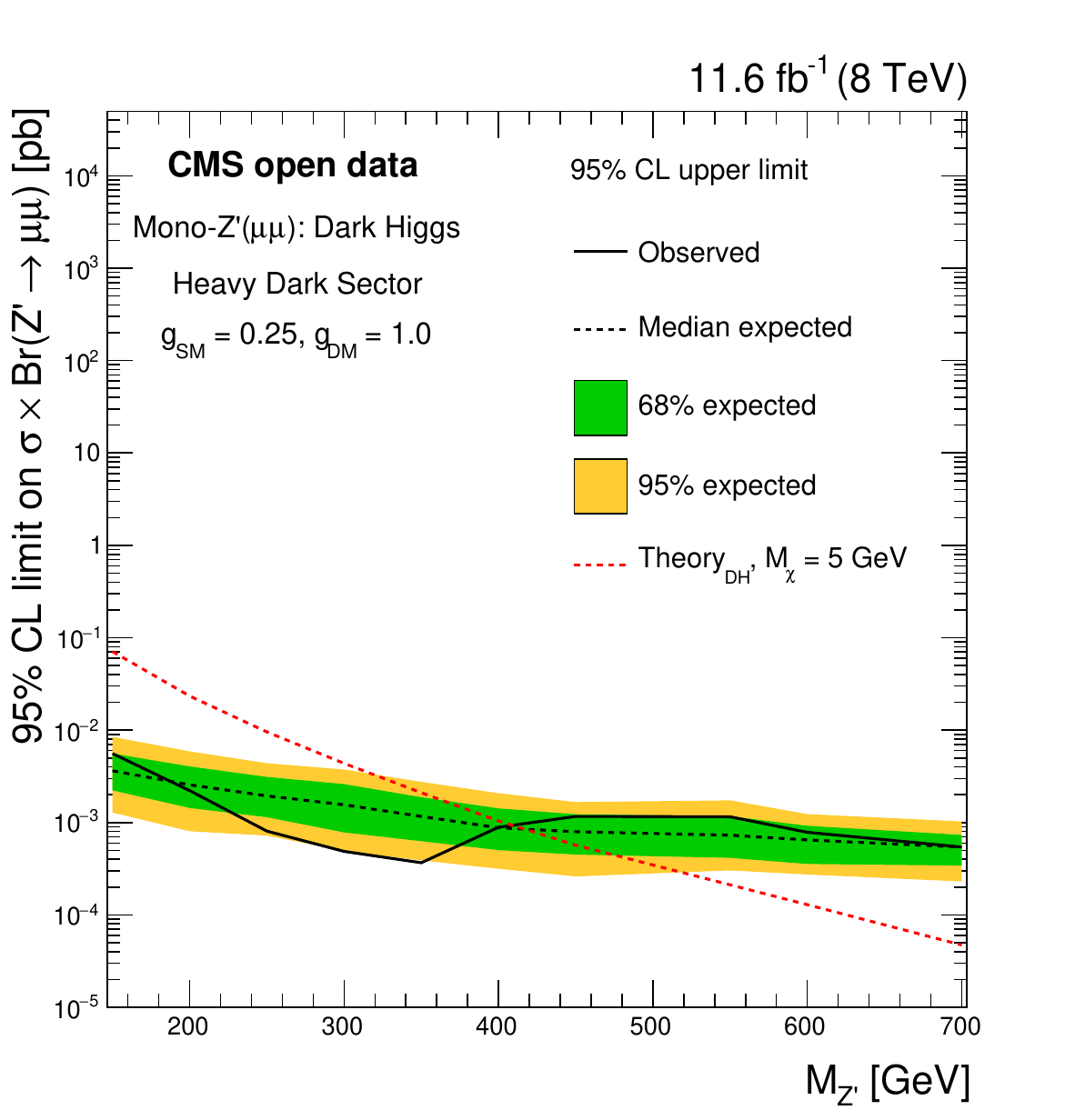}} 
  \caption{95\% CL upper limits on the cross section times the branching ratio (expected and observed), as a   function of the mediator's mass ($M_{Z^{\prime}}$), regarding the DH scenario, with the heavy dark sector set of masses and the muonic decay of the Z$^{\prime}$. 
  The red dotted line represents the dark Higgs model at a fixed dark matter mass ($M_{\chi} = 5$ GeV).}
  \label{figure:fig5}
\end{figure}

The cross section times the branching ratio $Br(Z' \rightarrow \mu\mu)$ limits for the simplified model (DH) is shown in figure \ref{figure:fig5}, with the heavy dark sector set of masses, the muonic decay of the Z$^{\prime}$ and coupling values of $g_{SM}$ = 0.25 and $g_{DM}$ = 1.0. The red dotted line represents the dark Higgs model at a fixed dark matter mass ($M_{\chi} = 5$ GeV).

\begin{figure}[h!]
\centering
  \resizebox*{8.cm}{!}{\includegraphics{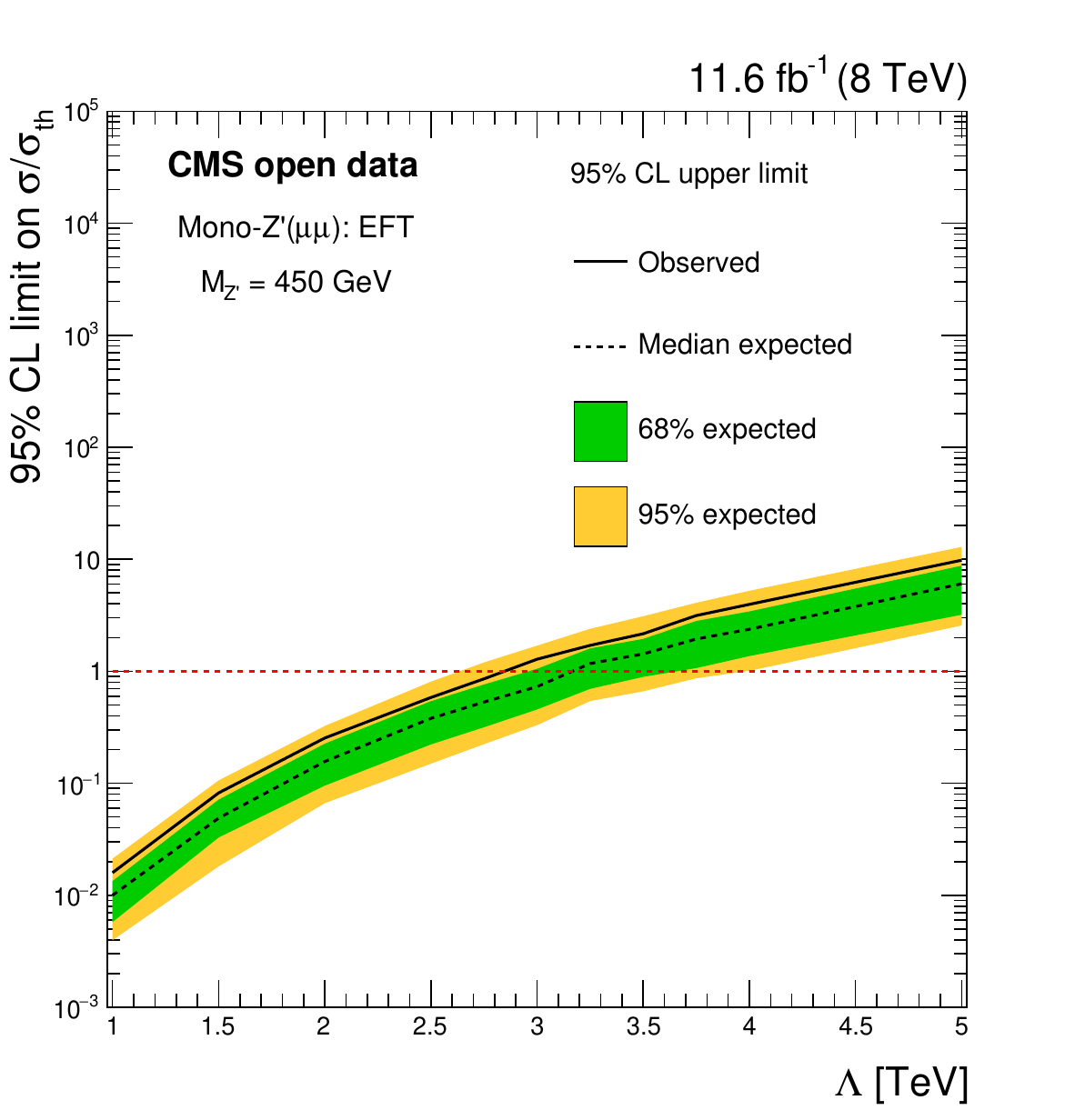}} 
  \caption{95\% CL upper limits on the signal strength (expected and observed), plotted against the cutoff scale of the EFT approach ($\Lambda$), at $M_{Z'}=$ 450 GeV. 
  The red horizontal dotted line represents the case $\sigma / \sigma_{\text{theory}}$ = 1.}
  \label{figure:fig7}
\end{figure}

The observed and expected upper limits at 95\% CL on the DM production cross section normalized to the predicted cross section, as a function of the model cutoff ($\Lambda$), is shown in figure \ref{figure:fig7} for the EFT scenario with $M_{Z^{\prime}} = 450$ GeV. 
The red horizontal dotted line represents the unity axis at which $\sigma = \sigma_{th}$.

Considering the DH scenario, we exclude the production of Z$^{\prime}$ with masses below 415 GeV for the expected median and 408 GeV for the observed data. 
Since the cross section measurements multiplied by the $ Z^{\prime} \rightarrow \mu^{+}\mu^{-}$ branching ratios for the DH scenario, which are listed in tables \ref{table:tabchi}, are varied only with the change of both of the mediator and dark higgs masses, and do not depend on the choice of dark matter mass \cite{R1}, so that the expected and observed limits on $M_{Z^{\prime}}$ will not change with the change of the DM mass.
For the EFT scenario, we also exclude the cutoff scale of the model below 3170 GeV for the expected median, and below 2850 GeV for the observed data. The introduced limits have been set at 95\% CL.
\section{Summary}
\label{section:Summary}
A study of the production of DM particles (WIMPs), at the CMS experiment has been performed. 
The production of DM particles alongside a new heavy gauge boson ($Z^{\prime}$) has been studied based 
on the Mono-$Z^{\prime}$ model. Two scenarios of the model were considered, which are; a simplified model so called the dark Higgs and the light vector with effective field theory coupling. 
The study has been performed for the muonic decay of the $Z^{\prime}$ boson.

In this analysis, we used the CMS open experimental data samples collected by the CMS detector for the proton-proton collisions at a centre of mass energy of 8 TeV in 2012 during Run-I of the LHC, which correspond to an integrated luminosity of 11.6 fb$^{-1}$.
The SM expected backgrounds for our signals events were built using the CMS open Monte Carlo samples generated by the CMS collaboration. 

The analysed data were founded in good agreement with the simulated SM backgrounds within the total uncertainty including the statistical and systematic components. Thus, no evidence for the existence of WIMPs was found.
95\% CL upper limits were set on some of the model free parameters; for the DH scenario with coupling values of $g_{SM}$ = 0.25, $g_{DM}$ = 1.0, the Z$^{\prime}$ boson masses below 415 GeV for the expected median and 408 GeV for the observed data have been excluded.  
Finally, the ranges below 3170 GeV for the expected median, and below 2850 GeV for the observed data, were also excluded for the cutoff scale ($\Lambda$) of the EFT scenario, with $M_{Z^{\prime}} = 450$ GeV.

\begin{acknowledgments}
Thanks and acknowledgments to the CMS collaboration for the open data project. Authors would like to thank Tongyan Lin, one of the authors of \cite{R1}, for sharing us the UFO for the Mono-Z$^\prime$ model, that were used for the generation of the model signals. We also thank Nicola De Filippis from the politecnico di Bari/INFN for allowing us to use the computing facilities to produce and hosting our ntuples at Bari tier 2 servers.

\end{acknowledgments}


\end{document}